# Driving action on the climate crisis through *Astronomers for Planet Earth* and beyond

Adam R. H. Stevens[1] and Vanessa A. Moss[2]

[1]International Centre for Radio Astronomy Research, The University of Western Australia, Crawley, WA 6009, Australia
[2]CSIRO Space and Astronomy, Marsfield, NSW 2122, Australia

**Summary**

While an astronomer's job is typically to look *out* from Earth, the seriousness of the climate crisis has meant a shift in many astronomers' focus. Astronomers are starting to consider how our resource requirements may contribute to this crisis and how we may better conduct our research in a more environmentally sustainable fashion. *Astronomers for Planet Earth* is an international organisation (more than 1,700 members from over 70 countries as of November 2022) that seeks to answer the call for sustainability to be at the heart of astronomers' practices.  In this article, we review the organisation's history, summarising the proactive, collaborative efforts and research into astronomy sustainability conducted by its members. We update the state of affairs with respect to the carbon footprint of astronomy research, noting an improvement in renewable energy powering supercomputing facilities in Australia, reducing that component of our footprint by a factor of 2–3.  We discuss how, despite accelerated changes made throughout the pandemic, we still must address the format of our meetings. Using recent annual meetings of the Australian and European astronomical societies as examples, we demonstrate that the more online-focussed a meeting is, the greater its attendance *and* the lower its emissions.

## 1. Introduction

The year 2019 was memorable for many reasons.  Not only was it the last year of pre-pandemic "normality", but it also saw the acceleration of the rise of astronomy sustainability. As an independent submission to the 2020 decadal survey of astronomy and astrophysics in the United States, Williamson, Rector & Lowenthal (2019) recommended that astronomers have "a collective impact model to better network and grow our efforts […] to address climate change" (p. 1). Similarly, in their white paper for the 2020 long-range plan for Canadian astronomy, Matzner et al. (2019) emphasise that "greenhouse gas emissions must be understood as significant research costs" (p. 1). Also in 2019, the Australian astronomical community conducted a mid-term review of its 2016–2025 decadal plan. Taking inspiration from the white papers of the US and Canadian communities, the first national-scale audit of the carbon footprint of astronomy was conducted as part of that review (Stevens et al., 2020). These papers collectively signalled the start of things to come. For example, sustainability was a focus in the 2021–2030 Strategic Plan for Astronomy in the Netherlands (Wijers, Kuijken & Wise, 2022).

The grass-roots organisation *Astronomers for Planet Earth* (A4E) was established in 2019 as a response by the astronomy community to the need to address the climate crisis (see overviews by Burtscher et al., 2021; Cantalloube et al., 2021; Frost et al., 2021; White et al., 2021a; or the independent article by Japelj, 2021). Two independent groups with an active interest in astronomy and sustainability came together to form A4E: one in North America (with a notable number of members from San Francisco State University) and one in Europe (with a similar-sized cluster of members from Leiden University). The organisation has since rapidly grown. With a membership base of more than 1,700 people in November 2022, we now have representation from over 70 countries (see Figure 1).

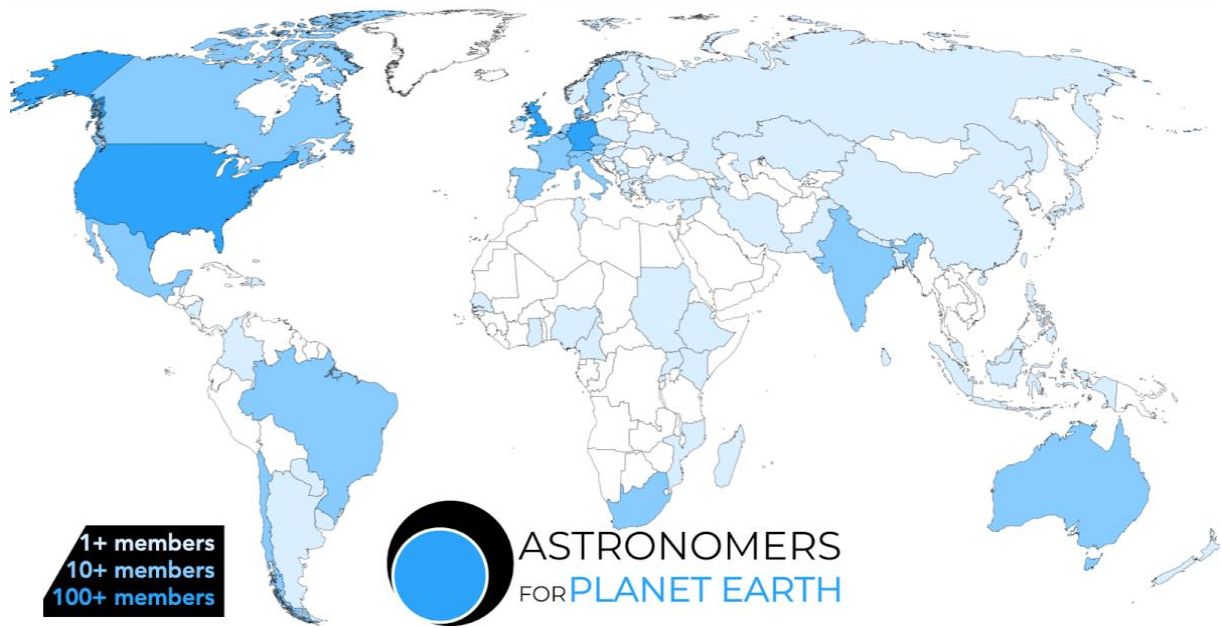

**Figure 1: Countries with members of *Astronomers for Planet Earth* as of November 2022. The shade of blue indicates whether the country has single-digit, tens, or hundreds of members.**

The mission statement of A4E is "To mobilise and empower the global astronomical community to take action on the climate crisis" (Cool, 2022). The urgency with which we must act to combat the climate crisis is clear from the latest International Panel on Climate Change assessment report (for a technical summary, see Pörtner et al., 2022). Physical scientists, like astronomers, hold an important position in this challenge, as we wield sufficient knowledge to understand the problem at hand, but are separated enough from its research such that when we press its importance, there cannot be claims of financial conflict of interest. A4E functions to empower astronomers with the resources necessary to engage the public on the climate crisis in their outreach. As astronomers, we can emphasise our perspective and understanding that Earth is the only place with life that we know of in the vast and ever-expanding cosmos.

This article presents an overview of how things have developed in the astronomy sustainability space over the three years since the inception of A4E. This includes the ever-increasing number of studies that have quantified the carbon emissions associated with astronomy and an outline of the efforts and recommendations of A4E members alongside other communities to help reduce these emissions. We pay particular attention to how future meetings may be conducted to better serve a sustainable future for astronomy.

## 2. The carbon emissions of astronomy

In one of the first emissions audit studies of astronomy, Stevens et al. (2020) considered four contributing factors to the carbon footprint of the job of an astronomer in Australia: air travel, operations of observatories, powering of supercomputing facilities, and office building electricity consumption. Based on an extrapolation from power-meter data supplied by the Pawsey Supercomputing Centre in Perth, Stevens et al. (2020) found supercomputing to be the single greatest source of carbon emissions, estimated at an average of 22 (confident in the range 14–28) tonnes of $CO_2$ emissions at ground level (t$CO_2$-e) per year (yr) per full-time-equivalent astronomer (FTE, including Masters and PhD students). The second biggest contributor was air travel, quoted to be 6 (5–7) t$CO_2$-e/yr/FTE, based on travel records from two institutes. As noted in the paper, however, this figure, based on airline quotes, did not account for non-$CO_2$ radiative forcing (e.g. from the production of contrails); other authors have interpreted the true value as double (i.e. 12 t$CO_2$-e/yr/FTE; Jahnke et al., 2020). Data from observatories placed a lower limit on their emissions of 4.5 t$CO_2$-e/yr/FTE (Stevens et al. accounted for most, but not all, observatories used by the Australian community) while powering offices sat at about 3 t$CO_2$-e/yr/FTE.

The figures quoted by Stevens et al. (2020) were based on data from 2018 and 2019. For most of 2020 and 2021, lockdowns, border closures, and grounded planes meant that the *air-travel* carbon footprint of Australian astronomers effectively dropped to zero. However, in 2022, work-related travel (both domestic and international) returned. While there remains to be a study that assesses whether the frequency of said travel has returned to 2019 rates, it is clear that the community has not fully embraced the opportunity to adapt to a more sustainable way of communicating and collaborating internationally, despite a broad range of emergent literature emphasising the positive benefits of doing so (e.g., Moss et al., 2021; Sarabipour et al., 2021).

However, the situation is not entirely bleak. Investments in renewable energy by the Australian Capital Territory government (see Mason, 2020) and Swinburne University of Technology (2020) now mean that two of the major supercomputing centres used by Australian astronomers are effectively carbon-neutral. We estimate this now places Australian astronomers' supercomputing carbon footprint at about 7 t$CO_2$-e/yr/FTE, a reduction of a factor of 2–3 since 2019. Air travel may have overtaken supercomputing as the largest source of Australian astronomers' emissions, emphasising how critical it is that we do not return to old flying habits. The greatest onus here is on senior staff members, as their average rate of flying exceeded postdocs by a factor of 3.8 and PhD students by a factor of 8.5, on average, prior to the COVID-19 pandemic (Stevens et al., 2020).

In a complementary study, Jahnke et al. (2020) calculated the carbon emissions of astronomers at the Max Planck Institute for Astronomy (MPIA) in Heidelberg, Germany. Compared to Stevens et al. (2020), Jahnke et al. found that, for the common sources of emissions considered in both studies, the job of being an astronomer was 2.7 times less carbon-intensive in Germany than in Australia. This reflects Australia's high reliance on coal burning for electricity production. Accounting for recent renewable investments in Australia suggests the difference is now closer to a factor of 1.6 (assuming the other quantities in Stevens et al., 2020, remain unchanged). Of course, renewable electricity bears no relevance to air-travel emissions. While air travel was evidently the greatest contributor to the average MPIA astronomer's emissions, this was 30% less than the average Australian astronomer. This could readily be explained by Australia's relative international isolation and its much lower population density, demanding longer average travel distances per trip.

Astronomers have since conducted subsequent carbon-audit studies in the Netherlands (van der Tak et al., 2021), France (Martin et al., 2022), and the United States (Simcoe et al., 2022). With the exception of van der Tak et al. (2021) — who do not consider observatories — the above studies all suggest astronomers carry a higher per-capita carbon footprint than the average citizen when controlling for location.

Individual observatories have also taken to quantifying their own operations' carbon footprint, including the Canada-France-Hawaii Telescope (Flagey et al., 2020) and the W. M. Keck Observatory (McCann et al., 2022). The European Southern Observatory (2022) has also calculated the emissions of ongoing operations of the entire organisation.

Many of these studies (sans Martin et al., 2022) have omitted at least two contributing factors of significance: (i) the emissions associated with *the construction* of observatories and (ii) the launch of space telescopes. Concerningly, Knödlseder et al. (2022) find that these two omissions potentially account for more than all other emissions sources of astronomers combined (though, as noted in the discussion of

Wilson (2022), estimating the emissions of space telescopes is challenging without raw data on the resource requirements of their construction, launch, and operation). Knödlseder et al. (2022) recommend that astronomers slow the construction rate of new infrastructure. Otherwise, any gains we may make through improving the sustainability of existing infrastructure will be outpaced by the additional requirements of new infrastructure.

**3. The future of meetings**

The COVID-19 pandemic forced global communication to shift online, accelerating pre-existing digital transformation trends. For astronomers and academics, this meant adapting how we run our major meetings, conferences, and workshops to work in a virtual space. Many people pointed out the opportunity this created to establish a new, permanent model for meetings that would vastly reduce carbon emissions due to a reduction in travel and increase accessibility to those without the means to travel (e.g., Stevens et al., 2020; Moss et al., 2021). Indeed, studies have calculated the carbon savings from online meetings (e.g., Burtscher et al., 2020; Tao et al., 2021), ranging from factors of tens to thousands lower than an equivalent in-person meeting.

Given that the global transition to effective online and hybrid events is ongoing, there have been many publications sharing learnings in this rapidly growing space, including the pros and cons of various meeting formats (e.g., Moss et al., 2020; Reshef et al., 2020; Cuk et al., 2021; Moss et al., 2021; Sarabipour et al., 2021; Skiles et al., 2021; Lowell et al., 2022; Moss et al., 2022). Alongside the positive impacts of virtual interaction on sustainability, there has also been extensive discussion on benefits in the context of accessibility and inclusivity, particularly when considering under-represented and less privileged groups in academia and our broader society (e.g. Sarabipour, 2020; Skiles et al., 2021; Köhler et al., 2022).

Similarly, various communities worldwide have become increasingly active in advocating for improved approaches to meeting formats. One example of this is *The Future of Meetings* (TFOM) community of practice, which formed following the 2020 same-named symposium (Moss et al., 2020, 2021). TFOM has grown since then from an organising committee predominantly consisting of Australian researchers to a global community actively advocating for improved widespread use of online/hybrid formats. The activities of TFOM have included providing advice and assistance to various conferences and meetings, sharing learnings in the form of publications and presentations, experimenting with new technologies in order to pioneer new approaches to online formats, and collaborating with diverse and varied groups on the theme of meetings and conferences (e.g., A4E, The University of Queensland's School of Information Technology and Electrical Engineering, SciConf historians, organisers of the 2024 International Astronomical Union (IAU) General Assembly, and Theo Murphy Initiative "Science for Public Good" workshop leads, among others). TFOM is just one example of the current global mobilisation for a better future of meetings and work, leveraging technology to improve accessibility, inclusivity, and sustainability in academia and broader society.

To demonstrate the significance of running astronomy conferences in an online/hybrid mode, we make a direct comparison between the emissions of three recent ASA (Astronomical Society of Australia) and EAS (European Astronomical Society) meetings in Figure 2. The 2020 and 2021 meetings were conducted under pandemic conditions, meaning no travel was possible. Despite this, the 2021 EAS and ASA meetings had their highest turnouts in history, with 2464 and 436 respective registrants (European Astronomical Society, 2021; The University of Melbourne, 2021). While the EAS meeting achieved this through a purely online meeting after a successful online model in 2020, the ASA meeting was conducted in a hybrid mode with local hubs in major cities throughout Australia for the first time. Of the 436 participants for the 2021 ASA meeting, 329 registered for their local hub, with the other 107 registered for online attendance.

One would expect the flight emissions of a purely in-person meeting to scale linearly with its number of participants (to first order). Indeed, that is consistent with a comparison of the 2019 ASA and EAS meetings. If the fraction of in-person attendees to a hybrid meeting remained constant, then we would still expect a positive correlation but with a shallower slope. Instead, we see that when major astronomy meetings are shifted to a hybrid model (ASA, 2022) and then to a fully online or distributed-hub model (ASA, 2021; EAS, 2020; EAS, 2021), the total emissions become reduced while the number of participants increases. We interpret this to mean that many attendees who would have attended an in-person meeting instead *choose* to attend online when given the option, and the number of extra people a hybrid or online meeting attracts significantly outweighs the number of people who *only* attend fully in-person meetings. Of course, to prove this statement applies generally, we should ideally have more data than are shown in Figure 2, which we defer to future work.

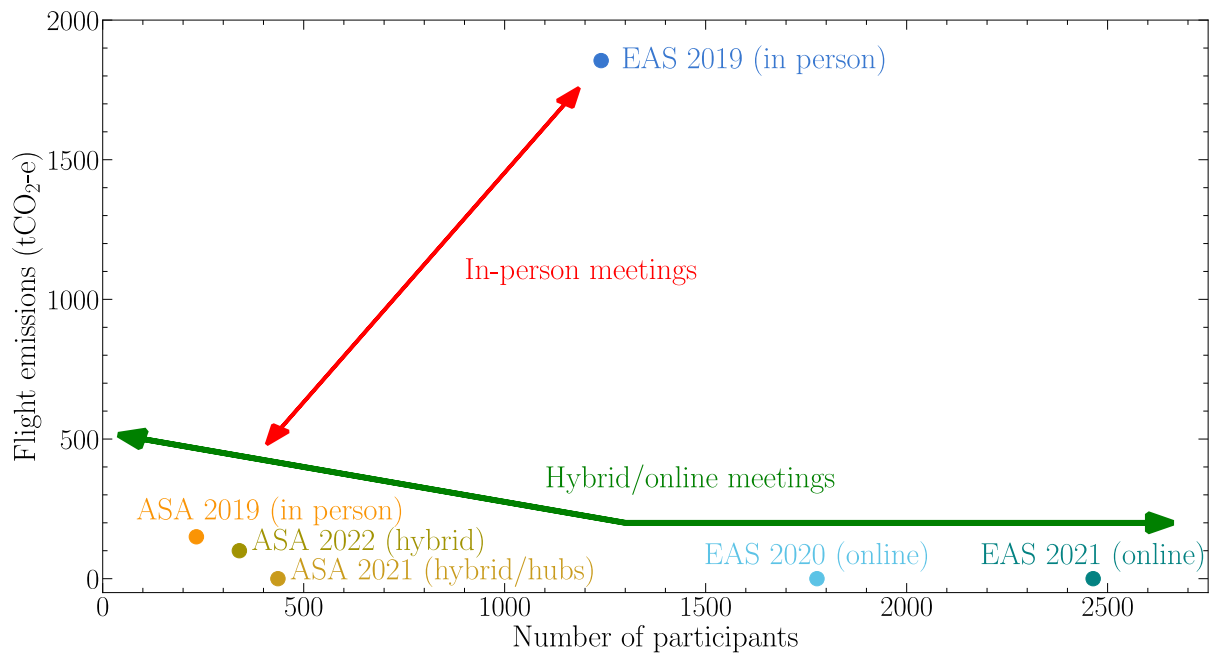

Figure 2: How the total flight emissions of six example major astronomy meetings have compared to their total number of participants and the mode of the meeting. Classical in-person meetings should scale linearly (red arrow). By contrast, the size of an online/hybrid meeting is anti-correlated with total emissions (saturating at zero; green arrow). This suggests that not only do many participants choose not to travel when given a viable option, but the meeting becomes increasingly accessible when there is a greater online focus. EAS 2019 figures were sourced from Burtscher et al. (2020). ASA figures were based on a list of cities of origin of participants, using Table 1 of Stevens et al. (2020) to calculate emissions, applying a factor of two for non-$CO_2$ radiative forcing, consistent with Barret (2020).

## 4. A4E Current Actions

### 4.1 Research

As an organisation of predominantly researchers, many A4E members actively research in the astronomy sustainability space. In September 2020, *Nature Astronomy* released a special *Climate Issue* (Nature Astronomy, 2020), with six articles published by A4E members on astronomy sustainability. Three of these articles included carbon audits, described in Section 2 (Flagey et al., 2020; Jahnke et al., 2020; Stevens et al., 2020), and a comparison of in-person versus virtual conferencing (Burtscher et al., 2020). In addition, Portegies Zwart (2020) discussed the carbon footprint of computation in astronomy — identified by Stevens et al. (2020) as one of the leading contributors to the field's overall footprint — highlighting the conflict between energy efficiency and time efficiency of some massively parallelised codes. Finally, Cantalloube et al. (2020) measured the potential effect of climate change on the seeing conditions at the Paranal Observatory in Chile, unsurprisingly finding that conditions were worsening.

Since then, there have been further independent studies of the effect of the changing climate at other observing sites (e.g. Haslebacher et al., 2022; van Kooten & Izett, 2022). The common conclusion among these works is that seeing conditions will gradually degrade, and the rate at which nights are lost due to bad weather will increase.

Research into astronomy sustainability has become sufficiently prevalent that *Nature Astronomy* now keeps a library of papers it has published in the area (Nature Astronomy, 2022). To date, this library includes >15 publications and two editorials.

4.2 Dissemination and communication

The A4E members have been central to establishing and running astronomy sustainability events at major national and international astronomy meetings. Special sessions were arranged at the fully virtual 2020 and 2021 EAS meetings. The respective sessions on "Astronomy for Future: Development, global citizenship & climate action" and "Astronomers for Planet Earth: Forging a sustainable future" included a range of talks and panel sessions from active researchers in astronomy sustainability and eminent scientists. With 100–300 participants each, these special sessions were among the best-visited sessions of the virtual EAS meetings (each with roughly 2,000 registrants across a dozen parallel sessions on average). A special session for the 2022 EAS meeting was also arranged, but not according to the original plan, as the meeting was designed to be almost exclusively in-person, despite advice from the EAS sustainability committee to the contrary (see Burtscher et al., 2022). Eventually, some hybrid aspects were added to this meeting (see Moss et al., 2022) by popular demand.

Special town hall sessions were also organised at the 2021 and 2022 annual ASA science meetings by A4E members (in the ASA's sustainability working group) on astronomy sustainability. The 2021 meeting also included a lively presentation that featured a celebrity scientist, titled "Astronomers! Your planet needs you", which is available on YouTube (Astronomical Society of Australia, 2021b). Further conferences and sessions by A4E members have included the 2021 conference of the Astronomical Society of the Pacific (Fischer et al., 2021) and the 2022 Chilean Astronomical Society meeting (Jaffé & Ramírez, 2022). Overview presentations of A4E have been given at multiple meetings of the American Astronomical Society (Sakari et al., 2020; White et al., 2021b) and at the CAP2021 conference (Frost et al., 2021).

A4E also runs a webinar series on YouTube (Astronomers for Planet Earth, n.d.), covering a wide range of topics related to astronomy sustainability. These include question-and-answer sessions between the invited speaker and A4E members.

In 2021, A4E released an open letter (Betancourt-Martinez et al., 2021) to astronomy institutes worldwide, calling on them to make sustainability a priority and to specifically impose carbon-reducing policies and communicate these changes to the community and public. Over 2,800 astronomers have signed this letter from more than 80 countries, well over double A4E's membership base at its time of release, equivalent to about 20% of the member base of the International Astronomical Union (IAU; IAU membership was not a requirement of signing the letter, nor does the IAU member base account for all astronomers). Several institutes have endorsed this letter, thereby making themselves accountable for the letter's request.

4.3 Bids for sustainable events

The A4E members have bid to host some of the largest astronomy meetings around the world in sustainable and inclusive formats, i.e. with an online focus. As recent examples, we proposed to host the 2027 General Assembly of the IAU and the 2023 Annual Science Meeting of the ASA, both with an online/hybrid focus that would have minimised the need for travel. Unfortunately, similar to the 2022 EAS meeting, the decision-makers for these events have actively pursued in-person-focused events instead. This is despite official statements from both the IAU (Lago & Christensen, 2021) and the Astronomical Society of Australia (2021a) supporting astronomers as they address climate change. Instead, A4E opted to run its own fully virtual symposium on astronomy sustainability and the broader aspects of the organisation in the week of 28 November – 2 December 2022, attracting more than 500 registrants (Wagner et al., 2023).

## 5. Conclusions

The movement of astronomy sustainability continues to grow, perhaps best exemplified by the continuous growth of the currently 1700-strong volunteer organisation *Astronomers for Planet Earth* (A4E). As physical scientists, it is important that astronomers use our position to lead by example in contributing to solutions to the climate crisis and in educating the wider public on it. Within A4E, we have researched and published carbon audits of the field of astronomy, worked with communities such as *The Future of Meetings* to outline how to conduct meetings with a hybrid/online focus that both reduce emissions and increase accessibility and attendance and conducted numerous workshops on astronomy sustainability. Further information and resources about the climate crisis are available on our website.*[1]

**Notes**

*[1] The Astronomers for Planet Earth website: https://astronomersforplanet.earth/

*[2] PYGAL: https://www.pygal.org/en/3.0.0/index.html

*[3] MATPLOTLIB: https://matplotlib.org/

**Acknowledgements**

We thank all A4E members for their contributions to the organisation but in particular the founding members and fellow members of the steering working group. We gratefully acknowledge comments on this manuscript from Leo Burtscher, Adrienne Cool, Andrea Gokus, Cenk Kayhan, Volker Ossenkopf-Okada, and Sarah Wagner.

Figures 1 and 2 in this paper were respectively prepared with the PYGAL*[2] and MATPLOTLIB*[3] packages for PYTHON.